\documentclass[twocolumn,prl,showpacs,preprintnumbers,
               superscriptaddress,floatfix]{revtex4}

\usepackage{amssymb}
\usepackage{graphicx}
\usepackage{dcolumn}
\usepackage{bm}
\usepackage{longtable}
\begin{document}

\title{A systematic study of the superdeformation of Pb isotopes with
relativistic mean field theory}

\author{Jian-You Guo}
\email{jianyou@ahu.edu.cn}
\affiliation{School of physics and
material science, Anhui university, Hefei 230039, P.R. China}
\author{Zong-Qiang Sheng}
\affiliation{Department of mathematics and physics, Anhui
University of Science and Technology, Huainan 232001, P.R. China}
\author{Xiang-Zheng Fang}
\affiliation{School of physics and material science, Anhui
university, Hefei 230039, P.R. China}

\begin{abstract}
The microscopically constrained relativistic mean field theory is
used to investigate the superdeformation for Pb isotopes. The
calculations show that there exists a clear superdeformed minimum
in the potential energy surfaces with four different interactions
NL3, PK1, TM1 and NLSH. The excitation energy, deformation and
depth of well in the superdeformed minimum are comparable for four
different interactions. Furthermore the trend for the change of
the superdeformed excitation energy with neutron number is
correctly reproduced. The calculated two-neutron separation energy
in the ground state and superdeformed minimum together with their
differences are in agreement with the data available. The larger
energy difference appearing in superdeformed minimum reflects a
lower average level density at superdeformations for Pb isotopes.
\end{abstract}

\pacs{21.10.-k, 21.60.Jz, 21.10.Dr}

\maketitle

Superdeformation(SD) of atomic nuclei is one of the most
interesting topics of nuclear structure studies. Over the past two
decades, many rotational bands associated with SD shapes have been
observed in several regions of the nuclear chart \cite{Singh},
with 85 SD bands observed in nuclei with $79<Z<84$ (the $A\sim
190$ region) alone, where an impressive number of results has been
obtained. Unfortunately, despite the rather large amount of
experimental information on SD bands, there are still a number of
very interesting properties, which have not yet been measured. The
characteristic examples are the spin, parity and excitation energy
relative to the ground state of the SD bands. The difficulty lies
with observing the very weak discrete transitions which link SD
levels with levels of normal deformation (ND levels). Until now,
only several SD bands have been identified to exist the
transitions from SD levels to ND levels in the $A\sim 190$ region:
two bands in $^{194}$Hg \cite{Khoo,Hackman}, and one band in each
of $^{194}$Pb \cite{Lopez,Hauschild}, $^{192}$Pb \cite{Wilson},
and $^{191}$Hg \cite{Siem}. Less precise measurements have been
achieved in $^{192}$Hg \cite{Lauritsen} and $^{195}$Pb
\cite{Johnson} following analysis of the quasi-continuum component
of the decay. Recently, the measurement of the excitation energy
of the yrast (lowest energy for a given spin) SD band in
$^{196}$Pb is reported \cite{Wilson05}, together with earlier
measurements of the excitation energies of SD states in $^{194}$Pb
and $^{192}$Pb, allows a systematic study of the energy of the SD
well in a single isotope chain. Many theoretical models have been
employed to study these superdeformed states of atomic nuclei. The
Strutinsky method with a Woods-Saxon potential \cite{Satula}, the
Hartree-Fock-Bogoliubov method with different mean field
parameterizations \cite{Krieger,Heenen,Libert}, and the cluster
model \cite{Adamian} have provided the predictions on the
excitation energy of SD bands, where a gross trend of decreasing
energy with decreasing neutron number is obtained for Pb isotopes,
but the absolute energies as well as their differences are not
consistently reproduced by these models as the analysis in
Ref.\cite{Wilson05}. Considering that the excitation energy and
the well depth of the SD minimum are amongst the most important
factors which affect the decay of the SD bands to the ground
state, the relativistic mean field (RMF) theory \cite{Serot86} has
also been recently applied to estimate the excitation energies and
depths of well for SD bands. The excitation energies and depths of
well in the SD minima of $^{194}$Hg and $^{194}$Pb have been
predicted in considerable agreement as compared with experiment
\cite{Lalaz98}. However, a systematic study of the energy of the
SD well in a single isotope chain presented in Ref.\cite{Wilson05}
with the nonrelativistic theories, has not been performed in a
relativistic framework. In resent years, the RMF theory has gained
considerable success in describing many nuclear phenomena for the
stable nuclei \cite{Reinhard89,Ring96} as well as nuclei even far
from stability \cite{meng98npa}. It has been shown that the RMF
theory can reproduce better the nuclear saturation properties (the
Coester line) in nuclear matter \cite{Brockmann90}, present a new
explanation for the neutron halo \cite{meng96prl} and predict a
new phenomenon --- giant neutron halos in heavy nuclei close to
the neutron drip line \cite{meng98prl}, give naturally the
spin-orbit potential, the origin of the pseudospin symmetry
\cite{Arima69,Hecht69} as a relativistic symmetry
\cite{Ginocchio97,meng99prc,guo05} and spin symmetry in the
anti-nucleon spectrum \cite{Zhou03prl}, and present good
description for the magnetic rotation \cite{Mad00}, the collective
multipole excitations \cite{Ma02}, the identical bands in
superdeformed nuclei \cite{konig93}, and the excitation energies
relative to the ground of SD bands \cite{Lalaz98}. Hence, here we
will report a systematic investigation of SD states for Pb
isotopes in the microscopic quadruple constrained RMF theory with
pairing treated by the BCS method, and show an excellent empirical
manifestations of this SD structure in Pb isotopes including the
evolution of the excitation energy, depth of well, deformation,
and shell structure as well as the comparison with the ND states.

The starting point of RMF theory is a standard Lagrangian density
where nucleons are described as Dirac particles which interact via
the exchange of various mesons including the isoscalar-scalar
$\sigma$ meson, the isoscalar-vector $\omega$ meson and the
isovector-vector $\rho$ meson. The effective Lagrangian density
considered is written in the form:
\begin{eqnarray}
\displaystyle
 {\cal L}
   & = &
     \bar\psi_i \left( i\rlap{/}\partial -M \right) \psi_i
    + \frac{1}{2} \partial_\mu \sigma \partial^\mu \sigma
    - U(\sigma)
    - g_{\sigma} \bar\psi_i \sigma \psi_i
   \nonumber \\
   &   & \mbox{}
    - \frac{1}{4} \Omega_{\mu\nu} \Omega^{\mu\nu}
    + \frac{1}{2} m_\omega^2 \omega_\mu \omega^\mu
    - g_{\omega} \bar\psi_i \rlap{/}{\mbox{\boldmath$\omega$}} \psi_i
   \nonumber \\
   &   & \mbox{}
    - \frac{1}{4} \vec{R}_{\mu\nu} \vec{R}^{\mu\nu}
    + \frac{1}{2} m_{\rho}^{2} \vec{\rho}_\mu \vec{\rho}^\mu
    - g_{\rho} \bar\psi_i \rlap{/} \vec{{\mbox{\boldmath$\rho$}}} \vec{\tau} \psi_i
   \nonumber \\
   &   &\mbox{}
    - \frac{1}{4} F_{\mu\nu} F^{\mu\nu}
    - e \bar\psi_i \frac{1-\tau_3}{2}\rlap{/}{\bf A} \psi_i ,
 \label{lagrangian}
\end{eqnarray}
where $\bar\psi=\psi^\dag\gamma^0$ and $\psi$ is the Dirac spinor.
Other symbols have their usual meanings.

The Dirac equation for the nucleons and the Klein-Gordon type
equations for the mesons and the photon are given by the
variational principle and can be solved by expanding the
wavefunctions in terms of the eigenfunctions of a deformed axially
symmetric harmonic-oscillator potential \cite{Gambhir:1990} or a
Woods-Saxon potential\cite{ZMR03}. The details can be also found
in Ref.~\cite{Ring96} and references therein.

The potential energy curve can be calculated microscopically by
the constrained RMF theory. The binding energy at certain
deformation value is obtained by constraining the quadruple moment
$\langle Q_2 \rangle$ to a given value $\mu_2$ in the expectation
value of the Hamiltonian~\cite{Ring80},
\begin{equation}
  \langle H'\rangle~
   =~\langle H\rangle
    +\displaystyle\frac{1}{2} C_{\mu} \left(\langle Q_2\rangle -\mu_2\right)^2,
\end{equation}
where $C_{\mu}$ is the constraint multiplier.

For the nuclei studied in this paper, the deformed harmonic
oscillator basis is taken into account and the convergence of the
numerical calculation on the binding energy and the deformation is
very good. The converged deformations corresponding to different
$\mu_2$ are not sensitive to the deformation parameter $\beta_0$
of the harmonic oscillator basis in a reasonable range due to the
large basis. The different choices of $\beta_0$ lead to different
iteration numbers of the self-consistent calculation and different
computational time. But physical quantities such as the binding
energy and the deformation change very little. Thus the
deformation parameter $\beta_0$ of the harmonic oscillator basis
is chosen near the expected deformation to obtain high accuracy
and low computation time cost. By varying $\mu_2$, the binding
energy at different deformation can be obtained. The pairing is
considered by the constant gap approximation (BCS) in which the
pairing gap is taken as $12/\sqrt{A}$ for even number nucleons.
\begin{figure}
\includegraphics[width=8.5cm]{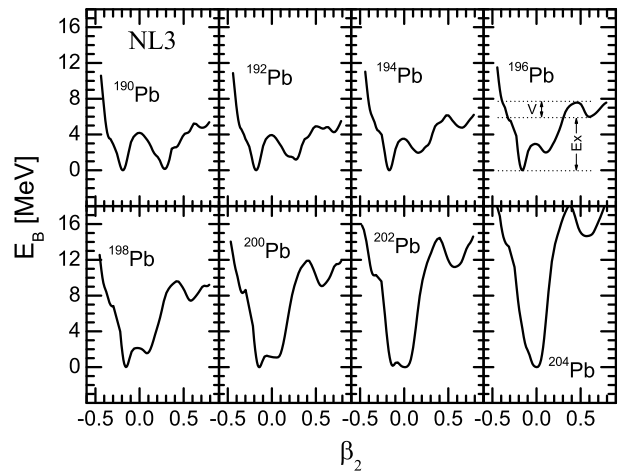}
\caption{The potential energy curves for $^{190-204}$Pb obtained
by the constrained RMF theory with the interactions NL3, where the
$E_x$ and $V$ represent respectively for the excitation energy
relative to the ground state of superdeformed minimum and the
depth of well of superdeformed minimum. The ground state binding
energy is taken as a reference. } \label{fig:pbnl3}
\end{figure}
\begin{figure}
\includegraphics[width=8.5cm]{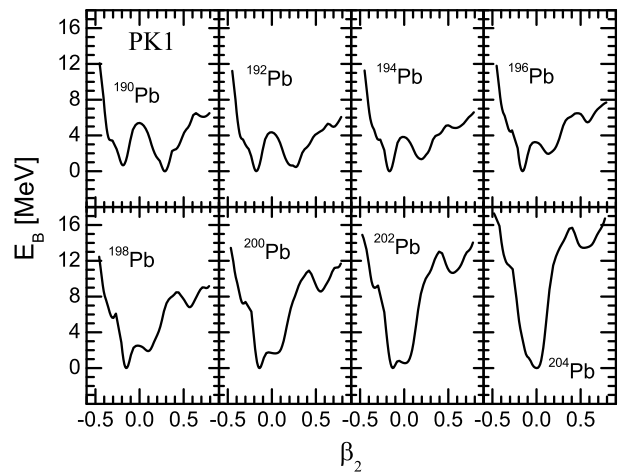}
\caption{The same as Fig.~\ref{fig:pbnl3}, but with PK1}
\label{fig:pbpk1}
\end{figure}
\begin{figure}
\includegraphics[width=8.5cm]{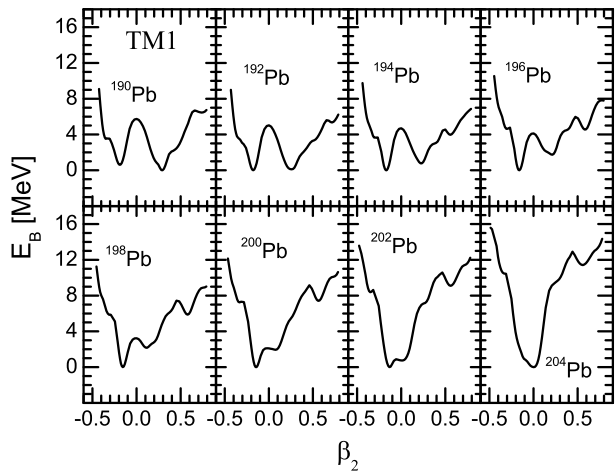}
\caption{The same as fig.~\ref{fig:pbnl3}, but with TM1}
\label{fig:pbtm1}
\end{figure}
\begin{figure}[tbp]
\includegraphics[width=8.5cm]{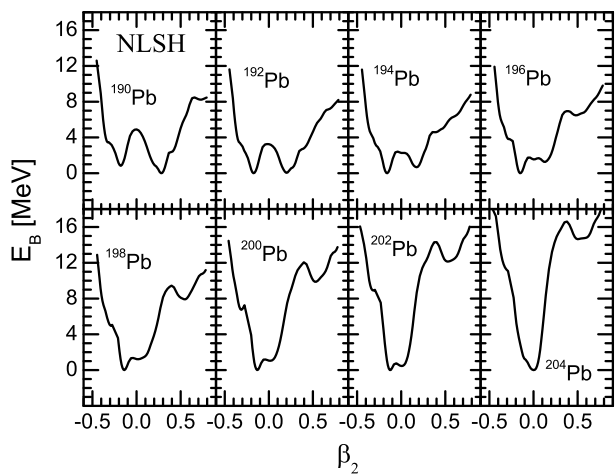}
\caption{The same as Fig.~\ref{fig:pbnl3}, but with NLSH}
\label{fig:pbnsh}
\end{figure}

The calculated potential energy curves for $^{190-204}$Pb are
exhibited respectively in the figures \ref{fig:pbnl3},
\ref{fig:pbpk1}, \ref{fig:pbtm1}, and \ref{fig:pbnsh} for the
interactions NL3\cite{LKR:1997}, PK1\cite{Long04},
TM1\cite{ST:1994} and NLSH\cite{SNR:1993}, in which the energy of
the ground state is taken as a reference. The $E_x$ and $V$ are
respectively the excitation energy and depth of well of the SD
minimum as shown in the subfigure for $^{196}$Pb. Similar patterns
are found for all the effective interactions. Most of the curves
display a clear SD minimum for the Pb isotopes, especial for the
nuclei with more neutron number. For $^{190,192}$Pb, the
calculated SD minimum is not very obvious, even disappears for the
interactions NLSH. For $^{190}$Pb, the RMF theory predicts a
considerable high excitation energy relative to the ground of SD
bands and shallow well in the SD minimum in comparison with its
neighboring nucleus $^{192}$Pb, which implies it is difficult to
come into being the stable SD state. For $^{192}$Pb, although the
well is shallow, the excitation energy is relatively lower, which
indicates that it is relatively easy to form the SD state as
observed in experiment. Start from the $^{194}$Pb, the RMF theory
predicts that the excitation energy increases with the increasing
of the neutron number in company with the increasing of depth of
well. So, the SD states can be still formed in these nuclei, which
agrees the experimental observations. However for the $N>118$
nuclei, as the excitation energy is too high to make it difficult
to excite the SD states. It may explain why only the SD nuclear
states between $N=110$ and $N=116$ are observed in the Pb isotope
chain. Besides the success in describing SD states, the RMF theory
predicts an interesting feature in the ground state. The evolution
of shape from the prolate to the oblate, and finally to the
spherical shapes are found in the Pb isotope chain. From Figs.1-4,
the ground state of $^{190}$Pb is exhibited a coexistence of the
prolate and the oblate with about 5 MeV stiff barrier against
deformation. Begin from $^{190}$Pb, the ground state gradually
moves toward the oblate side with smaller and smaller deformation
with the increasing of neutron number. Finally well spherical
$^{204}$Pb are seen.
\begin{figure}
\includegraphics[width=8.5cm]{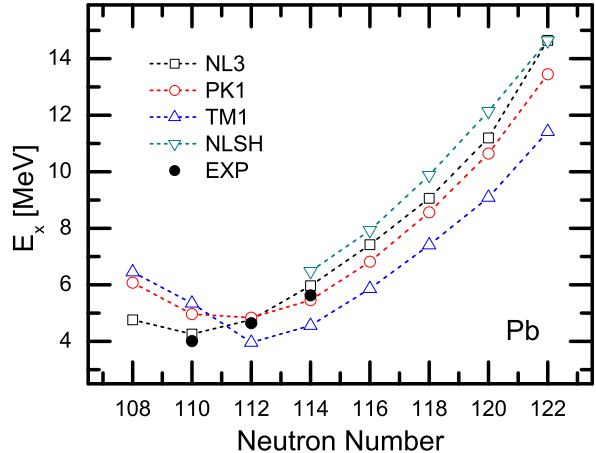}
\caption{ The SD bandhead energies of Pb isotopes $E_x$ as the
functions of neutron number obtained by the constrained RMF
calculations with the interactions NL3 (open squares), PK1 (open
circles), TM1 (up triangles) and NLSH (down triangles) in
comparison with the experimental data (filled circles).}
\label{fig:smnl1}
\end{figure}

In Fig.5, the calculated bandhead energies for SD bands as the
functions of neutron number are displayed for Pb isotopes, where
the open squares, open circles, up triangles, down triangles, and
filled circles stand respectively for the RMF calculations with
the interactions NL3, PK1, TM1, NLSH, and the data available
\cite{Wilson05}. First of all, trend for the change of the
excitation energies with neutron number are correctly reproduced
for all the effective interactions. Especially for the NL3, the
calculated excitation energies are in excellent agreement with the
data. The maximum deviation between theory and experiment is less
than 0.34MeV. For the TM1 interactions, the excitation energy is
overestimated for $^{190}$Pb and underestimated for
$^{192,194}$Pb. A jump appears in $^{194}$Pb, which disagrees the
experimental data. But the trend for the change of the excitation
energies with neutron number is in accordance with the RMF
predictions. For the PK1 interactions, except the excitation
energy of $^{192}$Pb is overestimated, theoretical predictions is
highly consistent with the experiment. For the NLSH interactions,
the RMF theory fails to reproduce the SD minimum for
$^{192,194}$Pb, but reproduce the SD minimum for $^{196}$Pb. In
addition, the trend for the change of excitation energies with the
neutron number is consistent with that in the other interactions.
Compared the RMF calculations, the Strutinsky method with a
Woods-Saxon potential \cite{Satula}, the Hartree-Fock-Bogoliubov
method with different mean field parameterizations
\cite{Krieger,Heenen,Libert}, and the cluster model\cite{Adamian}
predict only a gross trend of decreasing energy with decreasing
neutron number. The absolute energies and their differences are
not consistently reproduced by these model. It shows that the RMF
theory gives a better description of the SD excitation energies
for the Pb isotopes.
\begin{table}
\caption{\label{table:deform} The quadruple deformation $\beta_2$
and the depth of the superdeformed minimum $V$ in the
superdeformed states of $^{190-204}$Pb obtained by the constrained
RMF theory with the interactions NL3, PK1, TM1 and NLSH.}
\begin{ruledtabular}
\begin{tabular}{c|rrrr}
$\beta_2$   & NL3 & PK1 & TM1 & NLSH \\
\hline
 $^{190}$Pb&   0.71848&   0.71942&    0.73864&         \\
 $^{192}$Pb&   0.69986&   0.69862&    0.71772&         \\
 $^{194}$Pb&   0.65808&   0.57703&    0.55686&         \\
 $^{196}$Pb&   0.59984&   0.57800&    0.57885&   0.47840\\
 $^{198}$Pb&   0.58028&   0.56062&    0.57834&   0.53871\\
 $^{200}$Pb&   0.56225&   0.55835&    0.56068&   0.53811\\
 $^{202}$Pb&   0.56170&   0.54142&    0.55967&   0.52121\\
 $^{204}$Pb&   0.56161&   0.54071&    0.55935&   0.50177\\
\hline\hline
$V$   & NL3 & PK1 & TM1 & NLSH \\
\hline
$^{190}$Pb&   0.46439&   0.42181&   0.22936&     \\
$^{192}$Pb&   0.66217&   0.34837&   0.25777&     \\
$^{194}$Pb&   1.37963&   0.29612&   0.59873&     \\
$^{196}$Pb&   1.60176&   1.02856&   1.41779&   0.46892\\
$^{198}$Pb&   2.17652&   1.66011&   1.56866&   1.51712\\
$^{200}$Pb&   2.83555&   2.34135&   1.75199&   2.17761\\
$^{202}$Pb&   3.25852&   2.37082&   1.48711&   2.17321\\
$^{204}$Pb&   3.51464&   2.24769&   1.49029&   1.96709\\
\end{tabular}
\end{ruledtabular}
\end{table}

Besides the excitation energy, the deformation and well depth of
SD minimum are another two important parameters which reflect the
properties of the superdeformed states. In particular, the well
depth affects the life time of the superdeformed states. In Table
\ref{table:deform}, the quadruple deformation $\beta_2$ and the
depth of the superdeformed minimum $V$ in the superdeformed states
are listed respectively in the upper and lower panels for
$^{190-204}$Pb. Except for several exceptions, the deformation in
SD minima lies systemically between 0.5 and 0.7 for four different
interactions NL3, PK1, TM1 and NLSH, which agrees the observation
of superdeformed nuclei for excited states adopting ellipsoidal
shapes with an axis ratio around 2:1 \cite{Singh}. The RMF theory
predicts for the height of the barrier is lower than 1 MeV for
$^{190,192}$Pb, and higher than 1 MeV for $^{196-214}$Pb in all
the interactions. For $^{194}$Pb, the estimated barriers are
considerable different from different interactions.
\begin{table}
\caption{\label{table:sepaener} Two-neutron separation energy in
the ground state and superdeformation minimum obtained by the
constrained RMF theory with the interactions NL3, PK1, TM1 and
NLSH, in comparison with data available}
\begin{ruledtabular}
\begin{tabular}{c|rrrrr}
&&$S_{2n,{\rm ND}}$&[MeV]&\\
N   & NL3 & PK1 & TM1 & NLSH &EXP \\
\hline
110&    17.862& 16.915& 16.716& 17.984& 18.400\\
112&    17.537& 17.207& 17.135& 17.444& 17.810\\
114&    17.173& 16.885& 16.636& 17.255& 17.320\\
116&    16.739& 16.440& 16.124& 16.899& 16.820\\
118&    16.287& 16.050& 15.595& 16.652& 16.295\\
120&    15.709& 15.241& 14.886& 15.150& 15.837\\
122&    16.360& 15.287& 14.927& 15.611& 15.318\\
\hline\hline
&&$S_{2n,{\rm SD}}$&[MeV]&\\
N   & NL3 & PK1 & TM1 & NLSH   \\
\hline
110&    18.364& 18.025& 17.908& &\\
112&    17.031& 17.334& 18.436& &17.16(4)\\
114&    15.968& 16.266& 16.037& &16.31(4)\\
116&    15.287& 15.081& 14.820& 15.499&\\
118&    14.648& 14.303& 14.051& 14.653&\\
120&    13.574& 13.161& 13.202& 13.361&\\
122&    12.906& 12.484& 12.601& 12.637&\\
\end{tabular}
\end{ruledtabular}
\end{table}

Two-neutron separation energy defined as
$S_{2n}(Z,N)=E(Z,N)-E(Z,N-2)$ is sensitive quantity to test a
microscopic theory, where $E(Z,N)$ is the binding energy of
nucleus with proton number $Z$ and neutron number $N$. In Table
\ref{table:sepaener}, the two-neutron separation energies in the
ground state $S_{2n,{\rm ND}}$ and SD minimum $S_{2n,{\rm SD}}$
are shown respectively in upper and lower panels for the Pb
isotopes in comparison with the data\cite{Audi95,Wilson05}. From
there, it is found that the RMF calculations with four different
interactions well reproduce the experimental data for $S_{2n,{\rm
ND}}$. The maximum deviation between the calculations and data is
less than 1.7 MeV, especially for the NL3, the deviations is
within 1 MeV. For $S_{2n,{\rm SD}}$, the calculations with four
different interactions are comparable and close to the data
available. Both the calculations and experiment show that the
$S_{2n,{\rm ND}}$ and $S_{2n,{\rm ND}}$ vary smoothly with the
neutron number. No sharp drop in the binding energy is seen from
the $S_{2n}$, which indicates no significant shell gap appearing
in the Pb isotope chain whether the ground state or SD states.
\begin{table}
\caption{\label{table:edif}Two-neutron separation energy
difference in the ground state and superdeformation minimum
obtained by the constrained RMF theory with the interactions NL3,
PK1, TM1 and NLSH, in comparison with data available}
\begin{ruledtabular}
\begin{tabular}{c|rrrrr}
&&$\Delta S_{2n,{\rm ND}}$&[MeV]&\\
N   & NL3 & PK1 & TM1 & NLSH &EXP \\
\hline
110&   0.325&  -0.292&  -0.419&   0.540&   0.590\\
112&   0.364&   0.322&   0.499&   0.189&   0.490\\
114&   0.434&   0.445&   0.512&   0.356&   0.500\\
116&   0.452&   0.390&   0.529&   0.247&   0.525\\
118&   0.578&   0.809&   0.709&   1.502&   0.458\\
120&  -0.651&  -0.046&  -0.041&  -0.461&   0.519\\
\hline\hline
&&$\Delta S_{2n,{\rm SD}}$&[MeV]&\\
N   & NL3 & PK1 & TM1 & NLSH & EXP \\
\hline
110&   1.333&   0.691&   -0.528&  &\\
112&   1.063&   1.068&   2.399&   & 0.85(8)\\
114&   0.681&   1.185&   1.217&   & \\
116&   0.639&   0.778&   0.769&   0.846&\\
118&   1.074&   1.142&   0.849&   1.292&\\
120&   0.668&   0.677&   0.601&   0.724&\\
\end{tabular}
\end{ruledtabular}
\end{table}

In order to reveal further the detailed information on shell
structure, the two-neutron separation energy differences $\Delta
S_{2n}(Z,N)=S(Z,N)-S(Z,N+2)$ are presented in Table
\ref{table:edif}, where the $\Delta S_{2n,{\rm ND}}$ and $\Delta
S_{2n,{\rm SD}}$ represent respectively for those in the ground
state and superdeformed minimum with the data for
comparison\cite{Audi95,Wilson05}. From the Table \ref{table:edif},
it is found that the experimental $\Delta S_{2n,{\rm ND}}$ changes
typically around 0.5MeV, is well reproduced in the RMF
calculations for all the interactions with several exceptions.
Compared with the other interactions, the NL3 gives better
agreement with experiment. Only the deviation is relatively large
at $^{204}$Pb. Furthermore, the calculated $\Delta S_{2n,{\rm
ND}}$ shows very little differences for these nuclei with neutron
number from $N=112$ to $N=118$, is consistent with experiment
data, which suggests without a shell closure appearing at $N=112$
or $N=114$ predicted in other calculations. Compared with the ND
states, the SD separation energies $S_{2n}$ are significantly
larger than the typical ND value of 0.5 MeV, possibly reflecting a
lower average level density at superdeformations. In particular,
the $\Delta S_{2n,{\rm SD}}$ presents a obvious difference for
different nuclei. The $\Delta S_{2n,{\rm SD}}$ for these nuclei
with $N=112,114,118$ is much larger than that for their
neighboring nuclei, suggest a larger shell gap in these SD states
of $^{112,114,118}$Pb.

In summary, the superdeformation in $^{190-204}$Pb is investigated
by the microscopic quadruple constrained relativistic mean field
theory with all the most used interactions, i.e., NL3, PK1, TM1
and NLSH. The calculations show a clear SD minimum at nearly all
the potential energy curves for the Pb isotopes with similar
patterns for all the effective interactions. Trend for the change
of the excitation energies with neutron number are correctly
reproduced. The calculated deformation in SD minima lies
systemically between 0.5 and 0.7, is consistent with the
observation of experiment. The two-neutron separation energies in
the ground state and the SD minimum are well reproduced with
varying smoothly with the neutron number. Compared with the ND
states, the SD separation energies $S_{2n}$ are significantly
larger than the typical ND value of 0.5 MeV, possibly reflecting a
lower average level density at superdeformations.

\begin{acknowledgments}
This work was partly supported by the National Natural Science
Foundation of China under Grant No. 10475001, the Program for New
Century Excellent Talents in University of China, and the
Excellent Talents Foundation in University of Anhui Province in
China.
\end{acknowledgments}

\end{document}